\documentclass[twocolumn,prl,showpacs]{revtex4}

\usepackage{graphicx}
\usepackage{dcolumn}
\usepackage{bm}

\usepackage[english, activeacute]{babel}
\usepackage{rotating}
\usepackage{epsfig}
\usepackage{psfig}
\usepackage{amsmath}
\usepackage{amssymb}
\usepackage{amscd}
\usepackage{indentfirst}

\begin{document}


\title{Deformation of the N=Z nucleus $^{76}$Sr using 
       $\beta$-decay studies}

\author{E. N\'acher}\email{Enrique.Nacher@ific.uv.es} 
\author{A. Algora}\altaffiliation{MTA ATOMKI, H-4026
Debrecen, Hungary} \author{B. Rubio} \author{J.L. Ta\'{\i}n}
\author{D. Cano-Ott}\altaffiliation{CIEMAT, E-28040 Madrid,
Spain} 
\affiliation{Instituto de F\'{\i}sica Corpuscular, 
CSIC-Univ. de Valencia, E-46071 Valencia,Spain}

\author{S. Courtin} \author{Ph. Dessagne} \author{F. Mar\'echal}
\author{Ch. Mieh\'e} \author{E. Poirier}
\affiliation{Institut de Recherches Subatomiques, IN$_{2}$P$_{3}$-CNRS, 
F-67037 Strasbourg Cedex 2, France}

\author{M.J.G. Borge} \author{D. Escrig}
\author{A. Jungclaus}\altaffiliation{Univ. Autonoma de Madrid, E-28049 Madrid, Spain}
\author{P. Sarriguren} \author{O. Tengblad}
\affiliation{Instituto de Estructura de la Materia, CSIC,
E-28040 Madrid, Spain}

\author{W. Gelletly}\affiliation{Department of Physics, University of Surrey,
Guildford, GU2 5XH, United Kingdom}

\author{L.M. Fraile}\altaffiliation{Univ. Complutense de Madrid, E-28040 Madrid,
Spain} \author{G. Le Scornet}
\affiliation{ISOLDE, EP Division, CERN, CH-1211 Geneva, Switzerland}

\date{\today}

\begin{abstract}
A novel method of deducing the deformation of the N=Z nucleus $^{76}$Sr
is presented. It is
based on the comparison of the experimental Gamow-Teller strength 
distribution B(GT) from its $\beta$-decay with the results of QRPA calculations. 
This method confirms previous indications of the strong
prolate deformation of this nucleus in a totally independent way.
The measurement has been carried out with a 
large Total Absorption gamma Spectrometer, ``Lucrecia'', newly installed at
CERN-ISOLDE. 
\end{abstract}

\pacs{21.10.Pc, 23.40.-s, 23.40.Hc, 27.50.+e, 29.30.Kv, 29.40.Mc}%
\keywords{$^{76}$Sr, N=Z, Beta-decay, 
          Nuclear deformation, Total Absorption gamma Spectroscopy}%
\maketitle

The shape of the atomic nucleus is conceptually one of the simplest of its
macroscopic properties to visualise. However, it turns out to be one of the 
more difficult properties to measure. In general terms we now have a picture
of how the nuclear shape varies across the Segr\`e Chart. Nuclei near to the closed
shells are spherical. In contrast nuclei with the valence nucleons in between two
shells have deformed shapes with axial symmetry and the extent of the quadrupole 
deformation is
quite well described as being proportional to the product N$_{p}$N$_{n}$ of
the numbers of pairs of valence protons (N$_{p}$) and 
neutrons (N$_{n}$)~\cite{Casten:1985}. 
This picture is underpinned by both the Shell and Mean Field models of nuclear 
structure. Experiment and theory concur that, as the N$_{p}$N$_{n}$ 
parameterisation would suggest,
nuclei rapidly deform as we add only a small number of valence 
nucleons to the magic numbers. Thus 
nuclei in the middle of the f$_{7/2}$ shell turn out to be deformed even although
the numbers of valence nucleons are relatively small.\\\indent
Experimentally this 
picture is supported by a mass of independent 
observations: the strongly enhanced quadrupole transition rates between 
low-lying states, the 
strongly developed rotational bands built on low-lying states, and measurements 
of ground state quadrupole moments. Where we have evidence of the 
shapes of ground and excited states in the same nucleus they are, in general 
but not always, the same. It turns out that in some cases 
nuclear states with different shapes co-exist in the same 
nucleus~\cite{Andreyev:2000}.\\\indent
The nuclei with N$\approx$Z and A$\approx$70-80 are of particular interest in 
this context. Such nuclei enjoy a particular symmetry since the neutrons 
and protons are filling the same orbits. 
This, and the low single-particle level density, lead to
rapid changes in deformation
with the addition or subtraction of only a few nucleons. In terms of Mean Field 
models these rapid changes arise because of the proximity in energy of large 
energy gaps for protons and neutrons at Z,N=34 and 36 on the oblate and Z,N=38 
on the prolate side of the Nilsson diagram. As a result Mean Field calculations
predict the existence of several energy minima with quite different shapes in 
some of these nuclei~\cite{Nazarewicz:1985,Bonche:1985}. 
Evidence of this co-existence has been found for instance in Se and
Kr nuclei~\cite{Hamilton:1974,Chandler:1997}, and it is also predicted for the 
lightest Sr 
isotopes~\cite{Petrovici:1996}.
Thus it is of considerable interest to map out the deformation of 
the ground and excited states of nuclei in this region. This is easier said than
done, however. There are a number of methods to measure the deformation of the
ground state in unstable nuclei based on the interaction of the electric 
quadrupole moment of the nucleus with an external electric field 
gradient~\cite{Davni:1983,Hardeman:1991}. 
These techniques are not applicable to nuclei with J=0 or 1/2, moreover they 
very seldom give the sign of the quadrupole moment and hence cannot distinguish
between oblate and prolate shapes.\\\indent
Here we present an alternative method to deduce whether the ground state shape 
of an unstable nucleus is oblate or prolate, and apply it to the N=Z=38 nucleus
$^{76}$Sr, the most deformed nucleus in the region according to
Mean Field calculations~\cite{Bonche:1985} and previous in-beam
experiments~\cite{Lister:1990}.
The method is based on an accurate 
measurement of the Gamow-Teller strength distribution, B(GT), as a 
function of excitation energy in the daughter nucleus, and relies on the 
technique of Total Absorption gamma Spectroscopy (TAgS) which will be 
explained later. 
The theoretical idea was suggested by Hamamoto 
{\it et al.}~\cite{Hamamoto:1995} and pursued by Sarriguren 
{\it et al.}~\cite{Sarriguren:2001}.
According to them, one can study the deformation
of the ground state of a particular nucleus by measuring the
B(GT) distribution of its $\beta$-decay. In these references the authors
calculate the B(GT) distributions for various nuclei in the region
for the deformations minimising the ground state energy. In some cases,
the results differ markedly with the shape of the ground state of the 
parent, especially for the light Kr and Sr isotopes.\\\indent
A precise determination of the B(GT) distribution is required for such studies
and this is far from trivial. Traditional high resolution techniques, 
based on the use of high purity Germanium (HPGe) detectors to measure
the $\gamma$-rays emitted after the $\beta$-decay, often fail to
detect significant but very fragmented strength at high excitation 
energy in the
daughter nucleus. This is mainly due to three factors: the low 
photo-peak efficiency of HPGe 
detectors for high energy $\gamma$-rays, the high fragmentation of the B(GT) 
at high excitation energy, and the fragmentation of the gamma de-excitation 
of the levels in the daughter through many different gamma cascades.
Together they cause the so-called {\it Pandemonium effect}~\cite{Hardy:1977}:
many weak cascades de-exciting levels at high energy can remain
undetected leading to large systematic errors in the determination of the 
B(GT). This is the reason why, even although Refs.~\cite{Miehe:1997}
and~\cite{Dessagne:2004} give the 
first indication of the prolate character of the $^{76}$Sr ground state, 
one must determine the B(GT) distribution more accurately over the whole 
Q$_{EC}$ window to provide conclusive proof.\\\indent
The alternative, the TAgS technique, avoids 
these systematic uncertainties. The basis of this method is
the detection of the entire energy of the gamma cascades rather than individual 
$\gamma$-rays.
For this purpose one needs a high efficiency detector with 
acceptable resolution for gammas such as the
inorganic scintillators NaI(Tl) or BaF$_{2}$. Furthermore, this
detector must have a geometry as close as possible to 4$\pi$ to absorb the 
complete cascade energy. If this is achieved one can measure directly the 
$\beta$
intensity I$_{\beta}$(E) as a function of the excitation energy in the daughter
nucleus, and from this one can extract the B(GT) 
distribution.\\\indent
In this paper we present the results of TAgS measurements on the decay of the 
N=Z nucleus $^{76}$Sr. A comparison of the results of 
this measurement with the
calculations of Ref.~\cite{Sarriguren:2001} allows us to establish the prolate 
character of the ground state of $^{76}$Sr without ambiguity.\\\indent
With the aim of measuring the $\beta$-decay of nuclei far away from the
stability line with the Total Absorption technique, a spectrometer called 
``Lucrecia'' has been installed at the
ISOLDE mass separator at CERN. It consists of a large NaI(Tl) crystal of 
cylindrical shape (L=$\varnothing$=38 cm) with a cylindrical hole 
($\varnothing$=7.5 cm) at right angles to the symmetry 
axis. The purpose of the hole is twofold: on the one hand 
it allows the beam pipe
(coming from the separator) to enter up to the centre of the crystal,
thus allowing on-line activity of very short half-life ($>$5 ms) to be 
deposited here and measured. On the
other hand it allows us to place ancillary detectors inside for the 
detection 
of the positrons ($\beta^{+}$-decay), electrons ($\beta^{-}$-decay) or 
X-rays (EC process) produced in the 
decay.
Surrounding the whole setup there is a shielding box 
19 cm thick made of four layers: polyethylene-lead-copper-aluminium
(see Ref.~\cite{Poirier:2003} for further details).\\\indent 
In order to produce the nucleus
of interest ($^{76}$Sr), a \linebreak 52 g/cm$^{2}$ Nb target was 
bombarded with a 1.4 GeV proton beam.
The intensity was chosen to produce a counting
rate of $\approx$3.5 kHz in the NaI crystal.
In order to separate Sr selectively, a fluorination technique was 
used~\cite{Ravn:1975}. The radioactive beam was steered to the detector 
setup and implanted in an aluminised   
mylar tape which was moved every 15 seconds to transport the source to 
the middle of the crystal and
to avoid the buildup of the daughter activity
({\footnotesize T$_{ 1/2}$($^{76}$Sr)}=8.9 s, 
{\footnotesize T$_{ 1/2}$($^{76}$Rb)}=36.8 s).
During this 15 s cycle the decay of the implanted radioactive
source was measured. The $\gamma$-rays following the decay (either by 
$\beta^{+}$ or by EC) were measured by the NaI(Tl) crystal and analysed without
any condition on the ancillary detectors. However these detectors were very useful
for the on-line control of the measurement.\\\indent
In ideal conditions, if the TAgS had 100\% peak efficiency over the whole
energy range, the experimental spectrum measured in the NaI(Tl)
cylinder would be the $\beta$ intensity distribution 
I$_{\beta}$(E) convoluted with the energy 
resolution of the crystal and the response of the detector to the positron
when applicable. In reality the detector does not have 
100\% peak efficiency because of the dead material inside the spectrometer 
(the ancillary detectors) and the transverse hole. 
Consequently the spectrum is modified by the response
function of the detector. In other words, the
relationship between the quantity of interest, I$_{\beta}$(E), and
the experimental data d(i) is: 

\begin{equation}
d(i)=\sum_{j=1}^{j_{max}}R(i,j) \; I_{\beta}(j) \;\;\;\;\;\;\; 
{i \equiv channel \choose j \equiv energy~bin}
\label{eq:invmat}
\end{equation}

\begin{figure}[h,t]
  \includegraphics[width=.5\textwidth]{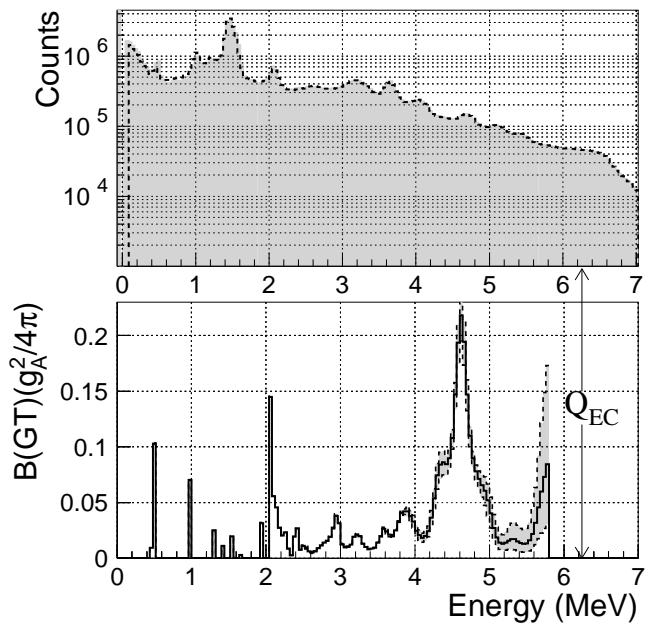}
  \caption{\label{fig:expresult}Upper panel: Experimental total absorption 
  spectrum of the $\beta$-decay of $^{76}$Sr overlaid with the recalculated 
  spectrum after the analysis (see text). Lower panel: B(GT) 
  distribution derived from the experimental data shown above. The shade
  represents the experimental uncertainty.}
\end{figure}

\noindent In order to obtain I$_{\beta}$(E) from our data we should solve 
Eq.~(\ref{eq:invmat}). This is not a trivial task because the response 
matrix $R(i,j)$ can not be inverted due to the fact that it is quasi-singular in the
sense that two neighbouring columns are very similar.
However, there is a set of algorithms that has been
developed to solve this kind of ``Ill Posed'' problem. In 
Ref.~\cite{Cano:2000} there is a systematic study of three of these methods
applied to the specific problem of the TAgS data. Here we have used the 
Expectation Maximisation algorithm~\cite{Dempster:1977} to obtain
the I$_{\beta}$(E) by unfolding the experimental data. To calculate the 
response matrix $R(i,j)$, which is needed by the algorithm,
we have used the levels and branching ratios given in Ref.~\cite{Dessagne:2004},
and the {\em GEANT4} simulation code. 
The analysis has been performed taking
into account both the EC and $\beta^+$ components of the decay.
A more detailed explanation on the procedure to calculate $R(i,j)$ and to analyse 
the data will be given in a forthcoming article.\\\indent
The best check one can perform to validate the result of the analysis is 
to recalculate the
experimental spectrum by multiplying the response function of the detector ($R(i,j)$
in Eq.~(\ref{eq:invmat})) by the resulting beta intensity I$_{\beta}$. If the 
analysis is
properly done, this recalculated spectrum should be very similar to the real 
experimental spectrum. The upper part of Fig.~\ref{fig:expresult} shows the
experimental spectrum (shade without line) overlaid with the recalculated one
(dashed line). The agreement between the two spectra is very good.\\\indent
The I$_{\beta}$(E) is the experimental result of this work, however, the physical
information is carried by the reduced transition probability B(GT), which can 
be extracted from the I$_{\beta}$(E) using the expression:

\begin{equation}
B(GT) = \frac{I_{\beta}(E)}{f(Q_{EC}-E)T_{1/2}}
           \times6147\left( \frac{g_{V}}{g_{A}} \right) ^{2}
\label{eq:bgt}
\end{equation}

\noindent where the B(GT) is averaged inside the 40 keV energy bin, and 
$f(Q_{EC}-E)$ is the Fermi integral which carries the information on both
the phase space available in 
the final state and the Coulomb interaction.
For the calculation of the B(GT) we have used the Q$_{EC}$ value 
from Ref.~\cite{Audi:2003,Sikler:2004}, the $T_{1/2}$ from~\cite{Dessagne:2004} and the
tabulated Fermi integral from~\cite{Nudat:1971}.\\\indent
In the lower panel of Fig.~\ref{fig:expresult} 
the resulting B(GT) distribution is presented. The analysis gives a total B(GT) 
of 3.8(6)g$_{A}^{2}$/4$\pi$ up to 5.6 MeV of which 
57\% is located in the resonance between 4 and 5 MeV. This resonance is weakly
visible in the almost structureless TAgS spectrum of the
upper panel. Its large B(GT) value is a consequence of the
strong dependence of the Fermi integral with the energy. At lower energy, the
marked B(GT) to levels at 0.5 MeV, 1.0 MeV and 2.1 MeV were already observed 
in~\cite{Dessagne:2004}, although the B(GT) values are in disagreement with our
results due to the already mentioned {\it Pandemonium effect}. 
It should be noted that the $\beta$-delayed proton emission 
(S$_p$=3.5 MeV) in this decay has been observed at excitation energies
from 4.8 to 5.8 MeV~\cite{Miehe:1997,Dessagne:2004}. However, the contribution 
of this component is very small, of the order of 2\% in B(GT) compared
to the decay through the $\beta$-delayed $\gamma$-rays studied here.

\begin{figure}[h,t]
  \includegraphics[width=.5\textwidth]{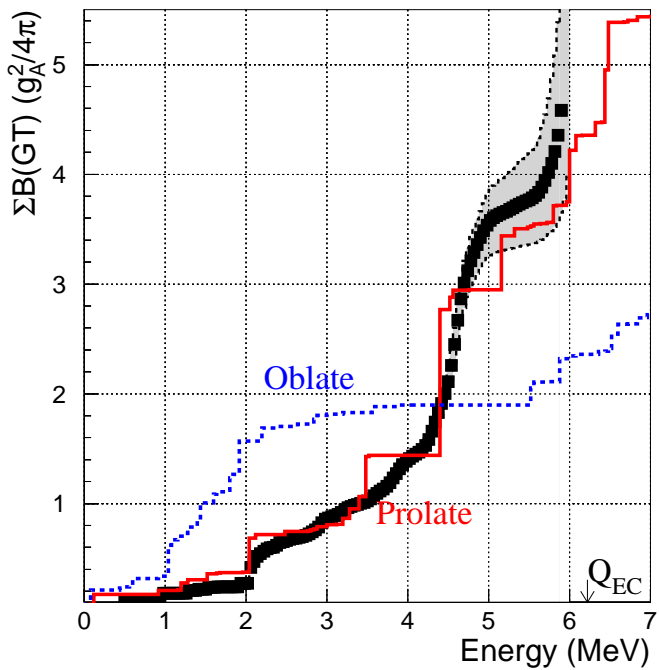}
  \caption{\label{fig:sumbgt}Accumulated B(GT) as a function of
  the excitation energy in the daughter nucleus. The experimental results 
  from this work (squares) are compared with the theoretical 
  calculations~\protect\cite{Sarriguren:2001}
  assuming prolate (solid line) and oblate (dashed line) shapes 
  for the $^{76}$Sr ground state.
  The shade indicates the experimental uncertainty.}
\end{figure}

One way to compare the results with the theory is to accumulate in each energy 
bin the sum of the B(GT) measured up to that particular energy.
Fig.~\ref{fig:sumbgt} shows this plot in which the experimental result
is compared with the theoretical calculations of  Ref.~\cite{Sarriguren:2001}
for both pure prolate and oblate shapes for the ground state of $^{76}$Sr. The
generally accepted quenching factor of 0.6 has been applied to the calculations.

In Ref.~\cite{Sarriguren:2001} the authors first construct 
the quasiparticle basis self-consistently from a deformed Hartree-Fock (HF)
calculation with density-dependent Skyrme forces and pairing correlations in
the BCS framework. 
The minima in the total HF energy vs deformation parameter plot give the possible 
deformations of the ground state. For the case of $^{76}$Sr two minima are found, 
one prolate with $\beta_{2}$=0.41 
and the other oblate with \linebreak $\beta_{2}$=-0.13. 
Finally, the QRPA equations
are solved with a separable residual interaction derived from the same Skyrme 
force used in the HF calculation. For the B(GT) calculation the same deformation is
assumed for the ground state of the parent and for the levels populated in the 
daughter nucleus. Fig.~\ref{fig:sumbgt} shows the
results using the residual interaction SK3. The agreement
of the experimental results of this work (squares in Fig.~\ref{fig:sumbgt})
with the prolate shape calculation 
of~\cite{Sarriguren:2001} is very good over the energy range 0-5.6 MeV. In 
contrast, there is no similarity between the results of the oblate 
calculation and the experimental points.\\\indent
This agrees with the strong deformation ($\beta_{2}$$\approx$0.4) 
of $^{76}$Sr already extracted from the dynamical properties observed in 
in-beam studies~\cite{Lister:1990}.
It also gives the first definitive experimental evidence of the prolate 
character of the ground state deformation, confirming the result
indicated in~\cite{Miehe:1997} and~\cite{Dessagne:2004}.\\\indent
We should point out here that these results prove the validity
of the method of deducing the sign of the electric quadrupole moment of ground
states or $\beta$-decay isomers from the study of their decay. This opens
new opportunities in the study of nuclei far from the stability line where very
often the first information comes from $\beta$-decay (half-life, 
J$^{\pi}$...).
On the other hand the theoretical approach used in the present study has been so
far restricted to nuclei in this region. The present work should encourage
further theoretical studies in other regions of well deformed nuclei.
It is worth noting here that QRPA calculations have been successfully applied to
describe other properties in nuclei with open shells where the nucleon pairing
correlations are important such as B(E2) values~\cite{Khan:2000} 
or giant resonances~\cite{Goriely:2002}.\\\indent
Finally, as a part of this series of experiments, we have recently published the
results on $^{74}$Kr decay~\cite{Poirier:2003}. In that paper a clear indication of
shape mixing was deduced, a conclusion which is further corroborated by the
present study which is free of shape admixtures.\\\indent 
Summarising, in this work we present the results of an experiment 
devoted to
measuring the B(GT) distribution in the decay of the N=Z isotope $^{76}$Sr. 
When we compare our experimental results with the theoretical calculations of 
Ref.~\cite{Sarriguren:2001} we conclude that the ground 
state of $^{76}$Sr is strongly prolate ($\beta_{2}$$\approx$0.4), in agreement 
with theoretical predictions~\cite{Bonche:1985,Petrovici:1996} 
and with previous experimental indications~\cite{Lister:1990,Miehe:1997,
Dessagne:2004}. An important consequence
of the present work is the validation of the method of deducing the deformation, 
including the sign of the quadrupole moment,
from the comparison of the $\beta$-decay TAgS results and the calculated B(GT) 
since the 
$^{76}$Sr ground state is a very clean case, free of shape admixtures.\\

The authors feel especially grateful to J.C. Caspar, J. Devin, G. Heitz and
C. Weber for all their invaluable work on the mechanical  
mounting of the spectrometer and the set-up of the data acquisition system. 
This work has been partially supported by the IN$_2$P$_3$/CNRS (France),
CICYT-MCyT (Spain), EPSRC (UK), the European Commission and the DFG (Germany).

\bibliography{76sr_prl_after_ref}

\end{document}